\documentstyle[12pt]{article}

\def\MS{{\overline{\hbox{MS}}}}
\def\tMS{{\overline{\hbox{2MS}}}}
\def\MSf{{\tiny\overline{\hbox{MS}}}}

\def\pd{\partial}
\def\G{\Gamma}
\def\Ga{\G}

\def\L{\Lambda}
\def\La{\L}
\def\m{\mu}
\def\l{\lambda}
\def\la{\l}
\def\s{\sigma}
\def\al{\alpha}

\def\va{\varphi}

\def\h{\hbox{${1\over2}$}}
\def\s{\hbox{${1\over6}$}}
\def\ha{\h}
\def\pa{\pd}
\def\ar{\rightarrow}
\def\bib{\bibitem}
\def\D{{\cal D}}
\def\dt{\frac{\hbar\la}{(4\pi)^{2}}\frac{\pa}{\pa t}}

\def\lo{(\hbox{\tiny{LL}})}
\def\nlo{(\hbox{\tiny{NLL}})}
\def\ol{(\hbox{\tiny{$1$-loop}})}

\def\be{\beta}

\def\ka{\kappa}

\def\beq{\begin{equation}}
\def\eeq{\end{equation}}
\def\bed{\begin{displaymath}}
\def\eed{\end{displaymath}}
\def\beqq{\begin{eqnarray}}
\def\eeqq{\end{eqnarray}}
\def\bedd{\begin{eqnarray*}}
\def\eedd{\end{eqnarray*}}

\advance\hoffset by -.35in
\begin{document}

\begin{flushright}
{\bf DIAS-STP 96-16\\September 1996}
\end{flushright}
\vspace{1cm}

\def\sqr#1#2{{\vcenter{\vbox{\hrule height.#2pt
        \hbox{\vrule width.#2pt height#1pt \kern#1pt
           \vrule width.#2pt}
        \hrule height.#2pt}}}}
\def\square{\mathchoice\sqr34\sqr34\sqr{2.1}3\sqr{1.5}3}

\begin{center}\Large{\bf{Resummation of the\\
Two Distinct Large Logarithms in the\\
Broken $O(N)$-symmetric $\phi^4$-model}}
\end{center}
\vspace{0.2cm}

\begin{center}
C. Wiesendanger\footnote{In collaboration with C. Ford,
Theoretisch-Physikalisches Institut,
Universit\"at Jena,\\
Fr\"obelsteig 1, D-07743 Jena, Germany}\\
{\it Dublin Institute for Advanced Studies\\
School of Theoretical Physics\\
10 Burlington Road, Dublin 4, Ireland}
\end{center}
\vspace{0.2cm}

\begin{abstract}
The loop-expansion of the effective potential in the
$O(N)$-symmetric $\phi^4$-model contains generically
two types of large logarithms. To resum those systematically
a new minimal two-scale subtraction scheme $\tMS$ is
introduced in an $O(N)$-invariant
generalization of $\MS$. As the $\tMS$ beta
functions depend on the renormalization scale-ratio
a large logarithms resummation is performed on them.
Two partial $\tMS$ renormalization group equations
are derived to turn the beta functions into $\tMS$
running parameters. With the use of standard perturbative
boundary conditions, which become applicable in $\tMS$,
the leading logarithmic $\tMS$ effective potential
is computed. The calculation indicates that there
is no stable vacuum in the broken phase of the theory
for $1<N\leq 4$.
\end{abstract}
\vspace{0.5cm}

\begin{center}Talk given at\\
{\it THE THIRD INTERNATIONAL CONFERENCE}\\
{\it RENORMALIZATION GROUP - 96}\\
Dubna, Russia, 26 - 31 August 1996\\
\vspace{0.25cm}
To be published in the proceedings
\end{center}
\clearpage

\section{Introduction}
\paragraph{}
There are many instances where an ordinary loop-wise
perturbative expansion is rendered useless by the
occurrence of large logarithms. This is the case eg. in
the discussion of scaling violation in deep
inelastic scattering (DIS) or in the determination of a
reliable approximation to the effective potential (EP) in
the standard model (SM). Only after resumming the large
logarithms does the violation of Bjorken scaling
yield one of the most accurate determinations of the
strong coupling constant \cite{dis1} or may the
requirement of vacuum stability be turned into sensible
bounds on the Higgs mass \cite{hig1}.

In the case of one type of large logarithms renormalization
group (RG) techniques are well established 
to perform the necessary resummation systematically.
However, in certain kinematical regimes in DIS there
are two types of large logarithms, in the SM EP for small
values of the Higgs field parameter there are five.
Although the problem has been recognized by many authors
no generally accepted RG techniques have been developed yet to
deal with those cases.

Sticking to the $\MS$ scheme the decoupling theorem
\cite{dec} was used in Ref. \cite{bando1} to obtain
some region-wise approximation to leading logarithms (LL) multi-scale
summations. Although this is perfectly reasonable, one
has to employ \lq\lq low-energy'' parameters, and it is
not clear  how to obtain sensible approximations for
these in terms of the basic parameters
of the full theory. Alternatively, one of us \cite{cf2}
argued that one could still apply the standard $\MS$ RG
equation to multi-scale problems provided \lq\lq improved''
boundary conditions were employed. Although such improved
boundary conditions were suggested in some simple cases,
no general prescription was given for constructing 
them, and no improved
boundary conditions were apparent for the subleading
logarithms summation. 

Clearly, one must go beyond the usual mass-independent
renormalization schemes if multi-scale problems are to
be seriously tackled. In the context of the EP we are
aware of two different approaches. In Ref. \cite{nak} it
was argued that one could employ a mass-dependent scheme
in which decoupling of heavy modes is manifest in the
perturbative RG functions. Alternatively, in Ref. \cite{ej}
the usual $\MS$ scheme was extended to include several
renormalization scales $\kappa_i$. While this seems to be an
excellent idea, the specific scheme in \cite{ej} has two drawbacks.
Firstly, the number of renormalization points does not
necessarily match the number of generic scales in the
problem at hand, as there is a RG scale $\ka_i$ associated
with each coupling. Secondly, when computing multi-scale RG
functions to $n$ loops one encounters contributions proportional
to $\log^{n-1}(\ka_i/\ka_j)$ (and lower powers). If some of the
$\log(\ka_i/\ka_j)$ are  large then even the perturbative RG
functions cannot be trusted and used to sum logarithms. A
similar approach to the one of Ref. \cite{ej} was outlined
in Ref. \cite{ni1} though no detailed perturbative
calculations were performed.

Here we outline a more systematic approach fully developed
to include next-to-leading logarithms (NLL) in Ref. \cite{fowi}.
In order to deal with the two-scale problem arising in
the analysis of the EP in the $O(N)$-symmetric
$\phi^4$-theory we introduce a $O(N)$-invariant
generalization of $\MS$. At
each order in a $\MS$ loop-expansion we perform a finite
renormalization to switch over to a new \lq\lq minimal two-scale
subtraction scheme'' $\tMS$ which allows for two renormalization scales
$\ka_i$ corresponding to the two generic scales in the problem.
The $\MS$ RG functions and $\MS$ RGE then split into two $\tMS$
\lq\lq partial'' RG functions and two \lq\lq partial'' RGE's.
The respective integrability condition inevitably imposes a
dependence of the partial RG functions on the renormalization
scale-ratio $\ka_2/\ka_1$. Supplementing the integrability
with an appropriate subsidiary condition we determine this
dependence to all orders in the scale-ratio and obtain a
trustworthy set of LL $\tMS$ RG functions. With the use of
the two \lq\lq partial'' RGE's we then turn those into LL
running two-scale parameters exhibiting features similar
to the $\MS$ couplings such as a Landau pole now in both
scaling channels. Using standard perturbative boundary conditions,
which become applicable in $\tMS$, we calculate the
effective potential in this scheme to LL and check it by
comparison with two-loop and next-to-large $N$ $\MS$
calculations. As a main result we find that for $1<N\le 4$
there is no stable vacuum in the broken phase. A full
analytic determination of the NLL corrections to the
results presented here is given in Ref.
\cite{fowi} and shows that the instability is not just an
artefact of a LL calculation.

\section{The one-loop effective potential in $\MS$}
\paragraph{}
Let us consider the $O(N)$-symmetric $\va^4$-theory with Lagrangian
\beq \label{1.1}
{\cal L}=\frac{1}{2}\pa_\al\va\;\pa^\al\va
-\frac{\la}{24}\va^4-\frac{1}{2}m^2\va^2-\La,
\eeq
where $\va$ is a real $N$-component scalar field. Note the inclusion of
the cosmological constant term which will prove essential in the
discussion of the RG and the effective potential later \cite{bando2}.

A loop-wise perturbation expansion of the effective potential
\cite{cw,ja} yields in the $\MS$-scheme to one loop
\beqq \label{1.3}
V^{(\hbox{\tiny{tree}})}\!&=&\!\frac{\la}{24}\va^4
+\frac{1}{2}m^2\va^2+\La, \nonumber \\
V^{\ol}\!&=&\!\frac{\hbar}{(4\pi)^{2}}
\frac{{{\cal M}_1}^2}{4}
\left(\log\frac{{\cal M}_1}{\mu^2}-\frac{3}{2}\right)
+\frac{\hbar}{(4\pi)^{2}}(N-1)\,\frac{{{\cal M}_2}^2}{4}
\left(\log\frac{{\cal M}_2}{\mu^2}-\frac{3}{2}\right),
\eeqq
where
\beq
{\cal M}_1=m^2+\h \la\va^2,\quad
{\cal M}_2=m^2+\s \la\va^2,
\eeq
and $\m$ is the $\MS$-renormalization scale.
The one-loop contribution to the EP thus contains logarithms of
the ratios ${\cal M}_i/\m^2$ to the first power and in general the
$n$-loop contribution will be a polynomial of the $n$th order
in these logarithms. (The explicit two-loop result has been
obtained in \cite{cf1}.)

In view of these logarithms the loop-wise expansion may be trusted only
in a region in field- and coupling-space where simultaneously
\beq
\frac{\hbar \la}{(4\pi)^2}\ll 1\quad\mbox{and}\quad
\frac{\hbar \la}{(4\pi)^2}
\log\frac{{\cal M}_i}{\mu^2}\ll 1.
\eeq
Due to the two largely differing scales ${\cal M}_i$ occurring in the
logarithms these conditions may hardly be fulfilled eg. around the tree-level
minimum of the potential, where ${\cal M}_2=0$, even with a judicious
choice of $\mu$. Hence, to obtain a sensible range of validity
one has to resum the logarithms in the EP.

In the one-scale case this would be achieved to LL by solving the
one-loop $\MS$ RG equation for the effective potential and by employing
the corresponding tree-level boundary conditions \cite{kast}. Here, we
have to deal with two relevant scales. The necessary generalization of
the $\MS$ scheme and the usual RG approach allowing for as
many renormalization scales as there are relevant scales in
the theory has been given in \cite{fowi}.

\section{The minimal two-scale subtraction scheme $\tMS$}
\paragraph{}
To track the two differing logarithms with two corresponding
renormalization scales we use the freedom of performing
a finite renormalization. Hence, to one loop we add a
\sl finite, \rm $O(N)$-\sl invariant \rm counterterm to the Lagrangian
\beq \label{2.1}
{\it\Delta}{\cal L}^{\ol}=\frac{\hbar}{(4\pi)^{2}}
\frac{{{\cal M}_1}^2}{4}\log\frac{{\mu}^2}{{\ka_1}^2}+
\frac{\hbar}{(4\pi)^{2}}
(N-1)\,\frac{{{\cal M}_2}^2}{4}\log\frac{{\mu}^2}{{\ka_2}^2},
\eeq
where the new renormalization scale $\ka_1$ is
tracking the Higgs logarithms and $\ka_2$ is tracking the
Goldstone logarithms.
Note that ${\it\Delta}{\cal L}^{\ol}$
is in fact a polynomial of fourth order in $\va$
consistent with renormalizability and the $O(N)$-symmetry.

In the minimal two-scale subtraction scheme
$\tMS$ thence introduced the
one-loop contribution to the EP becomes
\beq \label{2.2}
V^{\ol}=\frac{\hbar}{(4\pi)^{2}}
\frac{{{\cal M}_1}^2}{4}
\left(\log\frac{{\cal M}_1}{{\ka_1}^2}-\frac{3}{2}\right)
+\frac{\hbar}{(4\pi)^{2}}(N-1)\,\frac{{{\cal M}_2}^2}{4}
\left(\log\frac{{\cal M}_2}{{\ka_2}^2}-\frac{3}{2}\right).
\eeq
Hence, in $\tMS$ we may again trust the loop-expansion of
the EP at ${\ka_1}^2={\cal M}_1,\;{\ka_2}^2={\cal M}_2$ which
becomes the boundary condition for the RG evolution in the
two-scale case. Note that in this scheme the
beta functions inevitably depend on $\log(\ka_2/\ka_1)$
and will be trustworthy only after resummation of those
logarithms.

As discussed in detail in \cite{fowi} the general features
to be respected by $\tMS$ are:

i) The effective action $\Ga$, when expressed in terms of
the $\tMS$ parameters, should be independent of the
$\MS$ scale $\mu$.

ii) When $\ka_1=\ka_2$ $\tMS$ should
coincide with $\MS$ at that scale.

iii) When $N=1$ ($N\ar\infty$) the scale $\ka_2$ ($\ka_1$)
should drop and
$\tMS$ should coincide with $\MS$ at the remaining scale.

iv) When $\ka_i^2={\cal M}_i$ the standard loop-expansion
should render a reliable approximation to the full EP
insofar as ${\hbar\over {(4\pi)^2}}\la(\ka_1,\ka_2)$ is
\lq\lq small''.

Starting now from the identity
\beq \Ga_\MSf\bigl[\la_\MSf,m^2_\MSf,\La_\MSf,\va_\MSf;\mu\bigr]
=\Ga\bigl[\la,m^2,\La,\va;\ka_1,\ka_2\bigr]
\eeq
we derive the two partial $\tMS$ RGE's corresponding to variations of the
scales $\ka_i$, where the other scale $\ka_j$ and the $\MS$
parameters are held fixed, in much the same way as the $\MS$ RG
is usually derived. Specializing to the effective potential we obtain
\beq \label{2.15}
\D_i V=0,\quad
\D_i=\ka_i\frac{\pa}{\pa\ka_i}
+{_i\be}_{\la}\frac{\pa}{\pa \la}
+{_i\be}_{m^2}\frac{\pa}{\pa m^2}
+{_i\be}_\La\frac{\pa}{\pa\La}
-{_i\be}_\va\va\frac{\pa}{\pa\va}. \eeq
The two sets of RG functions are defined as usual
\beq \label{2.16}
{_i\be}_{\la}=\ka_i\frac{d \la}{d\ka_i},\quad
{_i\be}_{m^2}=\ka_i\frac{d m^2}{d\ka_i},\quad
{_i\be}_\La=\ka_i\frac{d\La}{d\ka_i},\quad
{_i\be}_\va\va=-\ka_i\frac{d\va}{d\ka_i}
\eeq
for $i=1,2$. In general they may be functions not only of
$\la,m^2$ as are the $\MS$ RG functions but also
of $\log(\ka_2/\ka_1)$.

Note that property ii) requires the sum of the
$\tMS$ RG functions at $\ka_1=\ka_2$ to coincide with
the $\MS$ RG function at that scale
\beq \label{2.17a}
_1\be_{.}(\ka_1=\ka_2)
+_2\be_{.}(\ka_1=\ka_2)
=\be_{.\,,\MSf},
\eeq
where the set of $\MS$ beta functions is given to one
loop by
\beqq \label{2.17b}
& &\be^{\ol}_{\la\,,\MSf}=\frac{\hbar}{(4\pi)^{2}}
\left(3+\frac{N-1}{3}\right){\la}^2,\quad
\be^{\ol}_{m^2\,,\MSf}=\frac{\hbar}{(4\pi)^{2}}
\left(1+\frac{N-1}{3}\right)\la m^2, \nonumber \\
& &\be^{\ol}_{\La\,,\MSf}=\frac{\hbar}{(4\pi)^{2}}
\left(\frac{1}{2}+\frac{N-1}{2}\right)m^4,\quad
\be^{\ol}_{\va\,,\MSf}=0.
\eeqq

In the $N=1$ limit property iii) fixes the $_1\be_{.}$
to be the usual $N=1$ $\MS$ RG functions, given to
${\cal O}(\hbar)$ by eqns. (\ref{2.17b}) with $N=1$, and requires
to disregard the second set of RG functions so that
$\D_2=\ka_2\pa/\pa\ka_2$. For $N\ar\infty$ 
there are no Higgs contributions and the $_2\be_{.}$
are the $N\ar\infty$ $\MS$ RG functions, again given to
${\cal O}(\hbar)$ by eqns. (\ref{2.17b}) 
in the large $N$ limit. The first set of RG functions
is then trivial, hence $\D_1=\ka_1\pa/\pa\ka_1$.

\section{The LL resummed $\tMS$ RG functions}
\paragraph{}
As we want to vary $\ka_1$ and $\ka_2$
independently we must respect the
integrability condition
\beq \label{2.18}
[\ka_1d/d\ka_1,\ka_2d/d\ka_2]=[\D_1,\D_2]=0,
\eeq
which allows us now to determine the $\tMS$ beta functions.
An essential feature of a mass-independent renormalization
scheme such as $\MS$ is that the beta functions do not
depend on the renormalization scale $\mu$. Unfortunately
we cannot generalize this to the multi-scale case
and demand that the two sets of beta functions be
independent of $\log(\ka_2/\ka_1)$. In fact, the independence
of the RG functions from the scales $\ka_i$, ie.
$[\ka_i\pd/\pd\ka_i,\D_j]=0$, is incompatible with the 
integrability condition (\ref{2.18}). However, as
we have one subsidiary condition at our disposal
it is possible to arrange eg. for the first set
of RG functions to be $\ka_i$-independent
\beq \label{2.19}
[\ka_i\pd/\pd\ka_i,\D_1]=0.
\eeq
Hence, at LL we have the first set of RG functions fixed
to be the $N=1$ values from eqns. (\ref{2.17b})
\beqq \label{2.20}
& &{_1\be}^{\lo}_{\la}=\frac{\hbar}{(4\pi)^{2}}3{\la}^2,\quad
{_1\be}^{\lo}_{m^2}=\frac{\hbar}{(4\pi)^{2}} \la m^2, \nonumber \\
& &{_1\be}^{\lo}_{\La}=\frac{\hbar}{(4\pi)^{2}}\frac{1}{2} m^4,\quad
{_1\be}^{\lo}_{\va}=0.
\eeqq

In general, we could assume a linear combination
${\tilde\be}_{.}=p\cdot{_1\be}_{.}+(1-p)\cdot{_2\be}_{.}$
of the two sets of beta functions to be $\ka_i$-independent.
As analyzed in detail in \cite{fowi} the results
for the beta functions, the running parameters and the
EP are then $p$-dependent. $p$ has to be fixed eg. by comparison
with the 2-loop and the next-to-large $N$ EP
and in our case it turns out that $p=1$
is the appropriate choice \cite{fowi}.

As $\D_1$ is now fixed eqn. (\ref{2.18}) yields RG-type equations
for the ${_2\be}_{.}$ which we solve next. Setting
\beq \label{3.1}
t=\frac{\hbar\la}{(4\pi)^{2}}\log\frac{\ka_2}{\ka_1}
\eeq
the equation for ${_2\be}_{\la}$ becomes to leading
order
\beq
-\dt\;{_2\be}^{\lo}_{\la}
+{_1\be}^{\lo}_{\la}\frac{\pa}{\pa \la}\;
{_2\be^{\lo}_{\la}}
-{_2\be}^{\lo}_{\la}\frac{\pa}{\pa \la}\;
{_1\be}^{\lo}_{\la}=0.
\eeq
The solution does not exlicitly depend on $t$
\beq
{_2\be}^{\lo}_{\la}(t)=\frac{\hbar}{(4\pi)^{2}}
\frac{N-1}{3}\la^2.
\eeq
Note that to fix the boundary conditions above and in what
follows we use property ii) leading to the relevant
condition (\ref{2.17a}).

We turn to the equation for ${_2\be}_{m^2}$
\beqq
& &-\dt\;{_2\be}^{\lo}_{m^2}
+{_1\be}^{\lo}_{\la}\frac{\pa}{\pa \la}\;
{_2\be^{\lo}_{m^2}}
-{_2\be}^{\lo}_{\la}\frac{\pa}{\pa \la}\;
{_1\be}^{\lo}_{m^2}=0.
\eeqq
This is easily solved by
\beq
{_2\be}^{\lo}_{m^2}(t)=\frac{\hbar}{(4\pi)^{2}}
\frac{N-1}{9}\left(1+2\;(1-3t)^{-1}\right)\la m^2.
\eeq

Next we determine ${_2\be}_{\La}$ from
\beqq
& &-\dt\;{_2\be}^{\lo}_{\La}
+{_1\be}^{\lo}_{\la}\frac{\pa}{\pa \la}\;
{_2\be^{\lo}_{\La}}
-{_2\be}^{\lo}_{\la}\frac{\pa}{\pa \la}\;
{_1\be}^{\lo}_{\La} \nonumber \\
& &\qquad+\;{_1\be}^{\lo}_{m^2}\frac{\pa}{\pa m^2}\;
{_2\be^{\lo}_{\La}}
-{_2\be}^{\lo}_{m^2}\frac{\pa}{\pa m^2}\;
{_1\be}^{\lo}_{\La}=0.
\eeqq
For later convenience we give the result partly in terms of
${_2\be}^{\lo}_{\la}(t)$ and ${_2\be}^{\lo}_{m^2}(t)$
\beq
{_2\be}^{\lo}_{\La}(t)=\frac{\hbar}{(4\pi)^{2}}
\frac{2(N-1)}{3}\;(1-3t)^{-\frac{2}{3}} m^4
+\frac{1}{2}\;{_2\be}^{\lo}_{\la}(t)\left(\frac{m^2}{\la}\right)^2
-{_2\be}^{\lo}_{m^2}(t)\frac{m^2}{\la}.
\eeq

Finally ${_2\be}_{\va}$ remains trivial
\beq
{_2\be}^{\lo}_{\va}(t)=0.
\eeq

It is obvious that \sl the beta functions possess Landau
poles \rm at $1-3t=0$. Hence, they are trustworthy only for $1\gg 3t$.
On the other hand, the limit $t\ar-\infty$ exists for the
whole set of ${_2\be}^{\lo}_{.}(t)$. This will allow
us later to discuss the non-trivial behaviour of the
two-scale EP around the tree-level minimum.

\section{The LL $\tMS$ running two-scale parameters}
\paragraph{}
The running parameters in $\tMS$
are functions of the variables
\beq \label{5.1}
s_i=\frac{\hbar}{(4\pi)^{2}}\log\frac{\ka_i(s_i)}{\ka_i},\quad
t=\frac{\hbar\la}{(4\pi)^{2}}\log\frac{\ka_2}{\ka_1},
\eeq
where $\ka_i$ are the reference scales. Note that $t(s_i)$ as
given in eqn. (\ref{3.1}) is in fact $s_i$-dependent,
$t=\frac{\hbar\la(s_i)}{(4\pi)^{2}}\log
\frac{\ka_2(s_i)}{\ka_1(s_i)}$. The above variables may
be expanded in series in $\hbar$ the LL terms of which
we determine now from eqn. (\ref{2.16}).

The equations for the leading order running two-scale
coupling are
\beq
\frac{d {\la}^{\lo}}{d s_1}=3\;{\la^{\lo}}^2,\quad
\frac{d {\la}^{\lo}}{d s_2}=\frac{N-1}{3}\;{\la^{\lo}}^2.
\eeq
They are easily integrated
\beq \label{4.42}
{\la}^{\lo}(s_i)=\la\left(1-3\la s_1
-\frac{(N-1)}{3}\la s_2\right)^{-1}
\eeq
with the boundary condition $\la(s_i=0)=\la$. Above, the
$s_1$-term accounts for the running of $\la$ due to the 'Higgs',
the $s_2$-term for the evolution due to the 'Goldstones'.

Next we determine the running mass from
\beq
\frac{d {m^2}^{\lo}}{d s_1}=\la^{\lo}\;{m^2}^{\lo}.
\eeq
This is easily solved
\beq \label{4.53}
{m^2}^{\lo}(s_i)=m^2\left(\frac{\la^{\lo}(s_i)}{\la}\right)
^{\frac{1}{3}}\;A(s_2).
\eeq
The constant of integration $A(s_2)$ is obtained
from the second $m^2$-equation
\beq \label{4.54}
\frac{d {m^2}^{\lo}}{d s_2}=
\frac{N-1}{9}\left(1+2\;\left(\frac{\la^{\lo}}{\la}
\left(1-\frac{(N+8)}{3}\;\la s_2-3t\right)\right)^{-1}\right)
\la^{\lo}\;{m^2}^{\lo}.
\eeq
We finally find
\beq
{m^2}^{\lo}(s_i)=m^2
\left(1-3\la s_1-\frac{(N-1)}{3}\la s_2\right)
^{-\frac{1}{3}}\left(\frac{1-\frac{(N+8)}{3}\;\la s_2-3t}{1-3t}\right)
^{-\frac{2}{3}\frac{N-1}{N+8}}.
\eeq
The boundary condition is chosen such that $m^2(s_i=0)=m^2$.

In order to obtain the running cosmological constant we
have to solve
\beq
\frac{d {\La}^{\lo}}{d s_1}=\frac{1}{2}\left({m^2}^{\lo}\right)^2.
\eeq
This yields the result
\beq \label{4.63}
{\La}^{\lo}(s_i)=\La-\frac{1}{2}
\left[\frac{\left({m^2}^{\lo}(s_i)\right)^2}{\la^{\lo}(s_i)}
-\frac{m^4}{\la}\right]
+B(s_2).
\eeq
To calculate the constant of integration $B(s_2)$ we
turn to the second $\La$-equation
\beqq \label{4.64}
\frac{d {\La}^{\lo}}{d s_2}\!&=&\!
\frac{2(N-1)}{3}\;\left(\frac{\la^{\lo}}{\la}
\left(1-\frac{(N+8)}{3}\;\la s_2-3t\right)\right)^{-\frac{2}{3}}
\left({m^2}^{\lo}\right)^2
\nonumber \\
\!&+&\!\frac{(4\pi)^{2}}{\hbar}\frac{1}{2}\;{_2\be}^{\lo}_{\la}
\left(\frac{{m^2}^{\lo}}{\la^{\lo}}\right)^2
-\frac{(4\pi)^{2}}{\hbar}\;{_2\be}^{\lo}_{m^2}\;
\frac{{m^2}^{\lo}}{\la^{\lo}}
\eeqq
and obtain the final result
\beqq \label{4.65}
{\La}^{\lo}(s_i)\!&=&\!-\frac{m^4}{2\la}
\left[\left(1-3\la s_1-\frac{N-1}{3}\la s_2\right)
^{\frac{1}{3}}\left(\frac{1-\frac{N+8}{3}\;\la s_2-3t}{1-3t}\right)
^{-\frac{4}{3}\frac{N-1}{N+8}}\;-\;1\right] \nonumber \\
\!&+&\!\frac{2 m^4}{\la}\frac{N-1}{N-4}(1-3t)^{\frac{1}{3}}
\left[\left(\frac{1-\frac{N+8}{3}\;\la s_2-3t}{1-3t}\right)
^{-\frac{N-4}{N+8}}\;-\;1\right]+\La.
\eeqq
Here the boundary condition is $\La(s_i=0)=\La$.
Due to the trivial ${_i\be}^{\lo}_{\va}$ the field parameter
$\va$ does not depend on $s_i$.

The LL running coupling $\la^{\lo}(s_i)$, and therefore
the running mass as well, have a Landau pole at
$1-3\la s_1-\frac{N-1}{3}\la s_2=0$ and clearly our
approximation will break down before this pole is 
reached. Of more importance is the behaviour of the
running cosmological constant as will be discussed at
the end of the next section.

\section{The LL RG improved $\tMS$ effective potential}
\paragraph{}
It is now an easy task to turn the results
for the running two-scale parameters into a RG improved effective
potential. $\D_i V=0$ yields the identity
\beq \label{6.1}
V(\la,m^2,\va,\La;\ka_1,\ka_2)=
V\bigl(\l(s_i),m^2(s_i),\va(s_i),\Lambda(s_i);\ka_1(s_1),\ka_2(s_2)),
\eeq
with $\ka_i(s_i)$ defined in (\ref{5.1}).
Next, we assume the validity of condition iv) from section 3.
Hence, if
\beq \label{6.2} \ka_i(s_i)^2={\cal M}_i(s_j)
\equiv m^2(s_j)+k_i\,\la(s_j)\va^2(s_j), \quad\quad
k_1={1\over 2},\;k_2={1\over 6} \eeq
the loop-expansion of the EP should render
a trustworthy approximation to the RHS of eqn. (\ref{6.1}).

To proceed we have to determine the values of
$s_i$ fulfilling (\ref{6.2}).
Insertion of the $\ka_i(s_i)$ from (\ref{6.2})
into (\ref{5.1}) yields a quite
implicit set of equations
\beq \label{6.3} s_i={\hbar\over{2(4\pi)^2}}
\log\frac{{\cal M}_i(s_j)}{\ka_i^2}. \eeq
However, since we are meant to be summing consistently leading
logarithms the explicit solution to this order is easily obtained
\beq
s_1^{\lo}={\hbar\over{2(4\pi)^2}}
\log\frac{m^2+\frac{1}{2}\,\la\va^2}{\ka_1^2},\quad
s_2^{\lo}={\hbar\over{2(4\pi)^2}}
\log\frac{m^2+\frac{1}{6}\,\la\va^2}{\ka_2^2}.
\eeq

At scales $s_i^{\lo}$ we can now approximate the
RHS of eqn. (\ref{6.1}) with the tree-level
contribution as displayed in (\ref{1.3}), hence
\beq \label{6.9}
V^{\lo}(\la,...;\ka_i)={\l^{\lo}(s_i^{\lo})\over{24}}\va^4
+{1\over2}{m^2}^{\lo}(s_i^{\lo})\va^2+\L^{\lo}(s_i^{\lo}).
\eeq
Insertion of the various expressions for the running
parameters yields the explicit, $O(N)$-invariant
final result for
the LL two-scale improved potential in $\tMS$ \cite{fowi}
\beqq \label{6.14}
V^{\lo}&=&{\la\va^4\over{24}}
\left(1-3\la s_1^{\lo}-\frac{N-1}{3}
\la s_2^{\lo}\right)^{-1} \nonumber \\
&+&{m^2\va^2\over2}
\left(1-3\la s_1^{\lo}-\frac{N-1}{3}
\la s_2^{\lo}\right)^{-\frac{1}{3}}
\left(1-{\frac{N+8}{3}\la s_2^{\lo}\over{1-3t}}\right)
^{-\frac{2}{3}\frac{N-1}{N+8}} \nonumber \\
&-&{m^4\over{2\la}}\left[
\left(1-3\la s_1^{\lo}-\frac{N-1}{3}
\la s_2^{\lo}\right)^{\frac{1}{3}}
\left(1-{\frac{N+8}{3}\la s_2^{\lo}\over{1-3t}}\right)
^{-\frac{4}{3}\frac{N-1}{N+8}}-1\right] \nonumber \\
&+&2\;{N-1\over{N-4}}\;{m^4\over{\la}}\;(1-3t)^{\frac{1}{3}}
\left[\left(1-{\frac{N+8}{3}\la s_2^{\lo}\over{1-3t}}\right)
^{-\frac{N-4}{N+8}}-1\right]+\La. \eeqq

There are various important checks on our result.
By construction it reduces in the single-scale
limits $N=1$ and $N\rightarrow\infty$ to the well-known one-scale
$\MS$ results \cite{kast}. A non-trivial check is provided by
expanding eqn. (\ref{6.14}) to second order in $s_i^{\lo}$.
As required the result of this expansion coincides with the leading
logarithmic terms in the explicit 2-loop effective potential as
obtained in Ref. \cite{cf1}. Furthermore, for $N\ar\infty$
we recover in the LL approximation the next-to-large $N$
expression for the EP as given in Ref. \cite{root}.
Finally, for $t=0$ the result (\ref{6.14}) has already been
obtained using the $\MS$ RG and a conjecture,
proven up to two loops, for the boundary condition
which becomes very involved in that approach \cite{cf2}.

We turn now to a discussion of the most important features of
the result (\ref{6.14}). In the broken phase ($m^2<0$)
the tree-level minimum is at ${\cal M}_2=0$ or
$s_2^{\lo}\ar-\infty$. Hence, as we approach it
$\log({\cal M}_2/{\cal M}_1)$ will become large.
If we are prepared to trust eqn. (\ref{6.14}) even
in the \sl extreme \rm case of the tree minimum
itself an intriguing property emerges.

As long as $N>4$ the $\va^4$-and $m^2\va^2$-terms vanish
and the $\La$-term converges to a finite value. As the slope
$\frac{dV^{\lo}}{ds_2^{\lo}}\tiny{(s_2^{\lo}\ar-\infty)}\searrow 0$
the EP takes its minimum in the broken phase at the
tree-level value and becomes complex for even smaller $\va^2$-values.

But for $1< N\leq4$ the $\La$-term, and thence $V^{\lo}$,
diverges to minus infinity indicating that for these
values of $N$ \sl there is no stable vacuum in
the broken phase. \rm Note especially that for $N=4$,
ie. the SM scalar boson content, there is still a
divergence. It is softer than for $N=2,3$, however,
as the penultimate term in eqn. (\ref{6.14})
becomes a logarithm
\beq
V^{\lo}=.....-{m^4\over{2\la}}(1-3t)^{\frac{1}{3}}
\log\left(1-{4\la s_2^{\lo}\over{1-3t}}\right)+\La.
\eeq

\section{Comment on NLL and Discussion}
\paragraph{}
The method presented in the calculation of the LL
two-scale effective potential is systematic.
In fact, in Ref. \cite{fowi} we have performed \sl a full analytic
computation \rm of the NLL two-scale RG functions, of
the corresponding NLL two-scale running parameters and
finally of the NLL effective potential $V^{\nlo}$.
Our main result is that for $1< N\leq4$ \sl the vacuum
instability in the broken phase persists \rm. Hence, it is
not a simple artefact of the LL resummation performed in this paper. 

The occurrence of a vacuum instability in the broken
phase of the $O(N)$-model raises immediately the
possibility of a similar outcome in a multi-scale analysis
of the SM effective potential. As the method outlined
generalizes naturally to problems with more than two
scales we are in a position to investigate systematically
the different possible scenarios. Because the SM analysis poses
a many-scale problem and will become
quite cumbersome it proves useful
to study first the effects of adding either fermions as in a
Yukawa-type model or gauging the simplest case of $N=2$
as in the Abelian-Higgs model. The Yukawa case is either a
two- or three-scale problem, depending on whether one includes
Goldstone bosons or not. The Abelian-Higgs model in
the Landau gauge will be a three-scale problem.
In the three-scale case one has \sl three \rm
integrability conditions $[{\cal D}_i,{\cal D}_j]=0$ and
three independent subsidiary conditions for free.
They are analogous to $[\ka_i\pa/\pa \ka_i,{\cal D}_1]=0$
which we used above. For the general $n$-scale problem
one would have $\ha n(n-1)$ integrability conditions
to be supplemented by $\ha n(n-1)$ subsidiary conditions.
The question of whether fermions or gauge fields may stabilize
the effective potential for small $N$
in a full multi-scale analysis is under investigation.

\section*{\bf Acknowledgments}
\paragraph{}
We express our warmest thanks to the organizers of RG96 for
realizing this stimulating meeting. This work has been partially
supported by Schweizerischer Nationalfonds.

\end{document}